\documentclass[twocolumn,preprint]{aastex631}

\usepackage{amsmath,amstext,natbib,apjfonts,longtable,verbatim,float,graphicx,color,longtable}

\newcommand{\jwst}{\textit{JWST}}
\newcommand{\CIII}{\hbox{{\rm C}\kern 0.1em{\sc iii}}}
\newcommand{\CIV}{\hbox{{\rm C}\kern 0.1em{\sc iv}}}

\shorttitle{CEERS: AGN Host Galaxies at z>3 with JWST}
\shortauthors{Kocevski et al.}

\begin{document}

\title{\large \bf CEERS Key Paper III: The Resolved Host Properties of AGN at 3 < z < 5 with \textit{JWST}}

\author[0000-0002-8360-3880]{Dale D. Kocevski}
\affiliation{Department of Physics and Astronomy, Colby College, Waterville, ME 04901, USA}

\author[0000-0002-0786-7307]{Guillermo Barro}
\affiliation{Department of Physics, University of the Pacific, Stockton, CA 90340 USA}

\author[0000-0001-8688-2443]{Elizabeth J.\ McGrath}
\affiliation{Department of Physics and Astronomy, Colby College, Waterville, ME 04901, USA}

\author[0000-0001-8519-1130]{Steven L. Finkelstein}
\affiliation{Department of Astronomy, The University of Texas at Austin, Austin, TX, USA}

\author[0000-0002-9921-9218]{Micaela B. Bagley}
\affiliation{Department of Astronomy, The University of Texas at Austin, Austin, TX, USA}

\author[0000-0001-7113-2738]{Henry C. Ferguson}
\affiliation{Space Telescope Science Institute, Baltimore, MD, USA}

\author[0000-0002-1590-0568]{Shardha Jogee}
\affiliation{Department of Astronomy, The University of Texas at Austin, Austin, TX, USA}

\author[0000-0001-8835-7722]{Guang Yang}
\affiliation{Kapteyn Astronomical Institute, University of Groningen, P.O. Box 800, 9700 AV Groningen, The Netherlands}
\affiliation{SRON Netherlands Institute for Space Research, Postbus 800, 9700 AV Groningen, The Netherlands}

\author[0000-0001-5414-5131]{Mark Dickinson}
\affiliation{NSF's National Optical-Infrared Astronomy Research Laboratory, 950 N. Cherry Ave., Tucson, AZ 85719, USA}

\author[0000-0001-6145-5090]{Nimish P. Hathi}
\affiliation{Space Telescope Science Institute, Baltimore, MD, USA}

\author[0000-0001-8534-7502]{Bren E. Backhaus}
\affiliation{Department of Physics, 196 Auditorium Road, Unit 3046, University of Connecticut, Storrs, CT 06269}

\author[0000-0002-5564-9873]{Eric F.\ Bell}
\affiliation{Department of Astronomy, University of Michigan, 1085 S. University Ave, Ann Arbor, MI 48109-1107, USA}

\author[0000-0003-0492-4924]{Laura Bisigello}
\affiliation{Dipartimento di Fisica e Astronomia "G.Galilei", Universit\'a di Padova, Via Marzolo 8, I-35131 Padova, Italy}
\affiliation{INAF--Osservatorio Astronomico di Padova, Vicolo dell'Osservatorio 5, I-35122, Padova, Italy}

\author[0000-0003-3441-903X]{V\'eronique Buat}
\affiliation{Aix Marseille Univ, CNRS, CNES, LAM Marseille, France}

\author[0000-0002-4193-2539]{Denis Burgarella}
\affiliation{Aix Marseille Univ, CNRS, CNES, LAM Marseille, France}

\author[0000-0002-0930-6466]{Caitlin M. Casey}
\affiliation{Department of Astronomy, The University of Texas at Austin, Austin, TX, USA}

\author[0000-0001-7151-009X]{Nikko J. Cleri}
\affiliation{Department of Physics and Astronomy, Texas A\&M University, College Station, TX, 77843-4242 USA}
\affiliation{George P.\ and Cynthia Woods Mitchell Institute for Fundamental Physics and Astronomy, Texas A\&M University, College Station, TX, 77843-4242 USA}

\author[0000-0003-1371-6019]{M. C. Cooper}
\affiliation{Department of Physics \& Astronomy, University of California, Irvine, 4129 Reines Hall, Irvine, CA 92697, USA}

\author[0000-0001-6820-0015]{Luca Costantin}
\affiliation{Centro de Astrobiolog\'ia (CSIC-INTA), Ctra de Ajalvir km 4, Torrej\'on de Ardoz, 28850, Madrid, Spain}

\author[0000-0002-5009-512X]{Darren Croton}
\affiliation{Centre for Astrophysics \& Supercomputing, Swinburne University of Technology, Hawthorn, VIC 3122, Australia}
\affiliation{ARC Centre of Excellence for All Sky Astrophysics in 3 Dimensions (ASTRO 3D)}

\author[0000-0002-3331-9590]{Emanuele Daddi}
\affiliation{Universit\'e Paris-Saclay, Universit\'e Paris Cit\'e, CEA, CNRS, AIM, 91191, Gif-sur-Yvette, France}

\author[0000-0003-3820-2823]{Adriano Fontana}
\affiliation{INAF - Osservatorio Astronomico di Roma, via di Frascati 33, 00078 Monte Porzio Catone, Italy}

\author[0000-0001-7201-5066]{Seiji Fujimoto}
\affiliation{Department of Astronomy, The University of Texas at Austin, Austin, TX 78712, USA}
\affiliation{Cosmic Dawn Center (DAWN), Jagtvej 128, DK2200 Copenhagen N, Denmark}
\affiliation{Niels Bohr Institute, University of Copenhagen, Lyngbyvej 2, DK2100 Copenhagen \O, Denmark}

\author[0000-0003-2098-9568]{Jonathan P. Gardner}
\affiliation{Astrophysics Science Division, Goddard Space Flight Center, Code 665, Greenbelt, MD 20771, USA}

\author[0000-0003-1530-8713]{Eric Gawiser}
\affiliation{Department of Physics and Astronomy, Rutgers, the State University of New Jersey, Piscataway, NJ 08854, USA}

\author[0000-0002-7831-8751]{Mauro Giavalisco}
\affiliation{University of Massachusetts Amherst, 710 North Pleasant Street, Amherst, MA 01003-9305, USA}

\author[0000-0002-5688-0663]{Andrea Grazian}
\affiliation{INAF--Osservatorio Astronomico di Padova, Vicolo dell'Osservatorio 5, I-35122, Padova, Italy}

\author[0000-0001-9440-8872]{Norman A. Grogin}
\affiliation{Space Telescope Science Institute, Baltimore, MD, USA}

\author[0000-0002-4162-6523]{Yuchen Guo}
\affiliation{Department of Astronomy, The University of Texas at Austin, Austin, TX, USA}

\author[0000-0002-7959-8783]{Pablo Arrabal Haro}
\affiliation{NSF's National Optical-Infrared Astronomy Research Laboratory, 950 N. Cherry Ave., Tucson, AZ 85719, USA}

\author[0000-0002-3301-3321]{Michaela Hirschmann}
\affiliation{Institute of Physics, Laboratory of Galaxy Evolution, Ecole Polytechnique Fédérale de Lausanne (EPFL), Observatoire de Sauverny, 1290 Versoix, Switzerland}

\author[0000-0002-4884-6756]{Benne W. Holwerda}
\affil{Physics \& Astronomy Department, University of Louisville, 40292 KY, Louisville, USA}

\author[0000-0002-1416-8483]{Marc Huertas-Company}
\affil{Instituto de Astrof\'isica de Canarias, La Laguna, Tenerife, Spain}
\affil{Universidad de la Laguna, La Laguna, Tenerife, Spain}
\affil{Universit\'e Paris-Cit\'e, LERMA - Observatoire de Paris, PSL, Paris, France}

\author[0000-0001-6251-4988]{Taylor A. Hutchison}
\affiliation{NSF Graduate Fellow}
\affiliation{Department of Physics and Astronomy, Texas A\&M University, College Station, TX, 77843-4242 USA}
\affiliation{George P.\ and Cynthia Woods Mitchell Institute for Fundamental Physics and Astronomy, Texas A\&M University, College Station, TX, 77843-4242 USA}

\author[0000-0001-9298-3523]{Kartheik G. Iyer}
\affiliation{Dunlap Institute for Astronomy \& Astrophysics, University of Toronto, Toronto, ON M5S 3H4, Canada}

\author{Brenda Jones}
\affiliation{Department of Physics and Astronomy, University of Maine, Orono, ME 04469-5709}

\author[0000-0002-0000-2394]{St{\'e}phanie Juneau}
\affiliation{NSF's NOIRLab, 950 N. Cherry Ave., Tucson, AZ 85719, USA}

\author[0000-0001-9187-3605]{Jeyhan S. Kartaltepe}
\affiliation{Laboratory for Multiwavelength Astrophysics, School of Physics and Astronomy, Rochester Institute of Technology, 84 Lomb Memorial Drive, Rochester, NY 14623, USA}

\author[0000-0001-8152-3943]{Lisa J. Kewley}
\affiliation{Center for Astrophysics | Harvard \& Smithsonian, 60 Garden Street, Cambridge, MA 02138, USA}

\author[0000-0002-5537-8110]{Allison Kirkpatrick}
\affiliation{Department of Physics and Astronomy, University of Kansas, Lawrence, KS 66045, USA}

\author[0000-0002-6610-2048]{Anton M. Koekemoer}
\affiliation{Space Telescope Science Institute, 3700 San Martin Dr., Baltimore, MD 21218, USA}

\author[0000-0002-8816-5146]{Peter Kurczynski}
\affiliation{Observational Cosmology Laboratory (Code 665), NASA Goddard Space Flight Center, Greenbelt, MD 20771, USA}

\author[0000-0002-9466-2763]{Aur{\'e}lien Le Bail}
\affil{Universit{\'e} Paris-Saclay, Université Paris Cit{\'e}, CEA, CNRS, AIM, 91191, Gif-sur-Yvette, France}

\author[0000-0002-7530-8857]{Arianna S. Long}
\affiliation{Department of Astronomy, The University of Texas at Austin, Austin, TX 78712, USA}

\author[0000-0003-3130-5643]{Jennifer M. Lotz}
\affiliation{Gemini Observatory/NSF's National Optical-Infrared Astronomy Research Laboratory, 950 N. Cherry Ave., Tucson, AZ 85719, USA}

\author[0000-0003-1581-7825]{Ray A. Lucas}
\affiliation{Space Telescope Science Institute, 3700 San Martin Drive, Baltimore, MD 21218, USA}

\author[0000-0001-7503-8482]{Casey Papovich}
\affiliation{Department of Physics and Astronomy, Texas A\&M University, College Station, TX, 77843-4242 USA}
\affiliation{George P.\ and Cynthia Woods Mitchell Institute for Fundamental Physics and Astronomy, Texas A\&M University, College Station, TX, 77843-4242 USA}

\author[0000-0001-8940-6768]{Laura Pentericci}
\affiliation{INAF - Osservatorio Astronomico di Roma, via di Frascati 33, 00078 Monte Porzio Catone, Italy}

\author[0000-0003-4528-5639]{Pablo G. P\'erez-Gonz\'alez}
\affiliation{Centro de Astrobiolog\'{\i}a (CAB), CSIC-INTA, Ctra. de Ajalvir km 4, Torrej\'on de Ardoz, E-28850, Madrid, Spain}

\author[0000-0003-3382-5941]{Nor Pirzkal}
\affiliation{ESA/AURA Space Telescope Science Institute}

\author[0000-0002-9946-4731]{Marc Rafelski}
\affiliation{Space Telescope Science Institute, 3700 San Martin Drive, Baltimore, MD 21218, USA}
\affiliation{Department of Physics and Astronomy, Johns Hopkins University, Baltimore, MD 21218, USA}

\author[0000-0002-5269-6527]{Swara Ravindranath}
\affiliation{Space Telescope Science Institute, 3700 San Martin Drive, Baltimore, MD 21218, USA}

\author[0000-0002-6748-6821]{Rachel S. Somerville}
\affiliation{Center for Computational Astrophysics, Flatiron Institute, 162 5th Avenue, New York, NY, 10010, USA}

\author[0000-0002-4772-7878]{Amber N. Straughn}
\affiliation{Astrophysics Science Division, NASA Goddard Space Flight Center, 8800 Greenbelt Rd, Greenbelt, MD 20771, USA}

\author[0000-0002-8224-4505]{Sandro Tacchella}
\affiliation{Kavli Institute for Cosmology, University of Cambridge, Madingley Road, Cambridge, CB3 0HA, UK}\affiliation{Cavendish Laboratory, University of Cambridge, 19 JJ Thomson Avenue, Cambridge, CB3 0HE, UK}
\author[0000-0002-1410-0470]{Jonathan R. Trump}
\affiliation{Department of Physics, 196 Auditorium Road, Unit 3046, University of Connecticut, Storrs, CT 06269, USA}

\author[0000-0003-3903-6935]{Stephen M.~Wilkins} %
\affiliation{Astronomy Centre, University of Sussex, Falmer, Brighton BN1 9QH, UK}
\affiliation{Institute of Space Sciences and Astronomy, University of Malta, Msida MSD 2080, Malta}

\author[0000-0003-3735-1931]{Stijn Wuyts}
\affiliation{Department of Physics, University of Bath, Claverton Down, Bath BA2 7AY, UK}

\author[0000-0003-3466-035X]{L. Y. Aaron\ Yung}
\affiliation{Astrophysics Science Division, NASA Goddard Space Flight Center, 8800 Greenbelt Rd, Greenbelt, MD 20771, USA}

\author[0000-0002-7051-1100]{Jorge A. Zavala}
\affiliation{National Astronomical Observatory of Japan, 2-21-1 Osawa, Mitaka, Tokyo 181-8588, Japan}

\begin{abstract}
We report on the host properties of five X-ray luminous Active Galactic Nuclei (AGN) identified at $3 < z < 5$ in the first epoch of imaging from the Cosmic Evolution Early Release Science Survey (CEERS).  Each galaxy has been imaged with the \textit{James Webb Space Telescope} (\jwst) Near-Infrared Camera (NIRCam), which provides spatially resolved, rest-frame optical morphologies at these redshifts.  We also derive stellar masses and star formation rates for each host galaxy by fitting its spectral energy distribution using a combination of galaxy and AGN templates.  The AGN hosts have an average stellar mass of ${\rm log}(M_{*}/{\rm M_{\odot}} )= 11.0$, making them among the most massive galaxies detected at this redshift range in the current CEERS pointings, even after accounting for nuclear light from the AGN.  We find that three of the AGN hosts have spheroidal morphologies, one is a bulge-dominated disk and one host is dominated by point-like emission. 
None are found to show strong morphological disturbances that might indicate a recent interaction or merger event.  Notably, all four of the resolved hosts have rest-frame optical colors consistent with a quenched or post-starburst stellar population.  The presence of AGN in passively evolving galaxies at $z>3$ is significant because a rapid feedback mechanism is required in most semi-analytic models and cosmological simulations to explain the growing population of massive quiescent galaxies observed at these redshifts.  Our findings are in general agreement with this picture and show that AGN can continue to inject energy into these systems after their star formation is curtailed, possibly helping to maintain their quiescent state.

\end{abstract}

\section{Introduction}

The role that Supermassive Black Holes (SMBHs) play in the evolution of galaxies remains a heavily debated topic within extragalactic astronomy.  There are signs that the growth of SMBHs and their host galaxies is closely connected, as evidenced by a variety of tight scaling relationships (e.g., \citealp{Magorrian98, Gebhardt00, Ferrarese00, McConnell13, Sun15}) and the need for energy injection in massive galaxies to limit their star formation activity \citep{Benson03, Croton06, Somerville08}.  As a result, Active Galactic Nuclei (AGN) have become key components in many galaxy evolution models (e.g., \citealp{Hirschmann12, Dubois16, Weinberger18, dave19, Zhu20,yung21}). However, several open issues remain in our understanding of how the SMBH-galaxy connection is established and maintained.  Among these are the mechanism(s) responsible for fueling the bulk of SMBH growth across cosmic time and the role of AGN in quenching the first generation of massive, quiescent galaxies.

The study of AGN host morphologies has widely been used to place constraints on the first of these issues.  Galaxy mergers are often invoked as a key process to potentially drive the co-evolution of galaxies and SMBHs due to their effectiveness at dissipating angular momentum and driving gas inflows that can both fuel black hole growth and build the stellar bulge via centrally concentrated starbursts (e.g., \citealp{Sanders88, kauffmann00, Springel05a, DiMatteo05, jogee06, Hopkins08a}). However, studies of X-ray selected AGN out to $z\sim2$ find that their host morphologies are no more disturbed than those of similar non-active galaxies 
\citep{Grogin05, Pierce07, Cisternas11, Schawinski11, Kocevski12, Villforth14, Rosario15}. This suggests other mechanisms such as minor mergers, Toomre-unstable ‘clumpy’ disks, or secular angular momentum loss play a larger role at fueling moderate-luminosity AGN at these redshifts than previously thought \citep{Hopkins14}.

A major caveat to these results is that SMBHs are predicted to accrete the bulk of their mass while heavily obscured and hydrodynamical simulations predict that this obscured phase should coincide with the most morphologically disturbed period of a galaxy interaction \citep{Hopkins06a}.  The morphologies of infrared-selected AGN and Compton-thick X-ray AGN do indeed show increased signs of disturbance relative to their unobscured counterparts, in apparent agreement with this scenario \citep{Koss10, Satyapal14, kocevski15, Donley18}.  At higher redshifts ($z>3$), however, the rest-frame optical morphologies of AGN have remained relatively unconstrained due to the fact that the reddest \emph{Hubble Space Telescope} (\emph{HST}) band, F160W, probes blueward of the Balmer break at these redshifts.

Another open question is whether AGN drive the quenching of star formation in massive galaxies.  Most cosmological models and simulations require a feedback mechanism to reproduce the properties of the massive galaxy population 
\citep{Springel05b, Croton06, Somerville08, Choi15, Weinberger18, dave19}, the needed energetics of which are not easily achieved by stellar feedback alone (e.g., \citealp{bower06}).  While AGN feedback has been implemented in a variety of ways (see \citealt{Somerville15} for a review), a common prescription in semi-analytic models involves major mergers that trigger AGN-driven winds that expel gas and eventually truncate the galaxy’s star formation activity \citep{Hopkins08a,Hopkins08b}.  

Although widely adopted, observational evidence for this scenario has remained elusive.   Demographic studies have produced mixed results, with findings of both negative and positive correlations between AGN activity and host properties such as star formation rates, rest-frame colors, and molecular gas content
\citep{Nandra07, Schawinski07, Mullaney12, Harrison12, Stanley15, Kirkpatrick19, Mountrichas22}.   However, there is a general consensus that X-ray selected AGN at $z\sim2$ are preferentially located in gas-rich, heavily star-forming galaxies (e.g., \citealt{Rosario15, florez20, florez21, Mountrichas21, Ji22}).  \citet{Ward22} point out that this is not necessarily in tension with the AGN feedback scenario given the common fuel supply that drives both AGN and star formation and the potential time delay between AGN activity and its effects.  

At higher redshifts ($z=3-5$), the discovery of a growing population of massive ($M_{*} > 10^{11}$ M$_\odot$) galaxies that fully quenched 1-2 Gyr after the Big Bang further necessitates a rapid and efficient quenching mechanism \citep{Schreiber18, Forrest20a, Forrest20b, Carnall22, Labbe22}.  However, the role that AGN play in this process is still uncertain, with some models finding that AGN feedback is not the leading mechanism shaping the bright end of the galaxy luminosity function at these redshifts \citep{Yung19a, Yung19b}.  If AGN are the drivers of early quenching among massive galaxies, we may see signatures of their impact on the properties of their host galaxies during this epoch. 

In this study, we provide a first look at the rest-frame optical morphology and star-formation activity of galaxies hosting X-ray selected AGN at $3<z<5$ using NIRCam imaging from the \emph{James Webb Space Telescope} (\emph{JWST}; \citealp{jwst}).  Our analysis is presented as follows. In Section 2 we describe the near-infrared imaging and X-ray data used for this study, while Section 3 describes our methodology for identifying AGN at our target redshift range. Section 4 describes our results, and the implications of our findings are discussed in Section 5.  When necessary the following cosmological parameters are used: $H_{0} = 70~{\rm km~s^{-1}~Mpc^{-1}; \Omega_{tot}, \Omega_{m}, \Omega_{\Lambda} = 1, 0.3, 0.7}$.

\section{Observations \& Data Description}

\subsection{CEERS Data}

The Cosmic Evolution Early Release Science Survey (CEERS) is an early release science program that will cover 100 arcmin$^{2}$ of the Extended Groth Strip (EGS) with imaging and spectroscopy using coordinated, overlapping parallel observations by most of the \emph{JWST} instrument suite.  CEERS is based around a mosaic of 10 NIRCam pointings, with six NIRSpec and six MIRI pointings observed in parallel.  Here we make use of the first four CEERS NIRCam pointings (hereafter epoch 1), obtained on 21 June 2022, known as CEERS1, CEERS2, CEERS3, and CEERS6.  In each NIRCam pointing, data were obtained in the short-wavelength (SW) channel F115W, F150W, and F200W filters, and long-wavelength (LW) channel F277W, F356W, F410M, and F444W filters.  The total exposure time for pixels observed in all three dithers was typically 2835 s per filter.

\begin{deluxetable*}{lcccccc}[t]
\tablenum{1}
\tablecolumns{8}
\tablecaption{Properties of our primary sample of X-ray detected AGN at $3<z<5$ \label{tbl:sample}}
\tablehead{
 \colhead{AEGIS-XD ID} & \colhead{R.A.} & \colhead{Dec} & 
 \colhead{z} & \colhead{z Type} & \colhead{log ($M_{*}/{\rm M_{\odot}})$} & 
 \colhead{$L_{\rm X, 0.5-10~keV}$} \vspace{-0.1in} \\ 
 \colhead{} & \colhead{(J2000)} & \colhead{(J2000)} & 
 \colhead{} & \colhead{} & \colhead{} & \colhead{($\times10^{44}$ erg s$^{-1}$)} }
\startdata
 AEGIS 482 & 214.755245 & 52.836807 &  3.465 & spec & 9.79$^{+0.08}_{-0.04}$ & 3.73$^{+0.31}_{-0.29}$ \\ 
 AEGIS 495 & 214.871261 & 52.845092 &  3.54$^{+0.05}_{-0.03}$ & phot & 11.01$^{+0.10}_{-0.02}$ & 0.18$^{+0.11}_{-0.10}$ \\ 
 AEGIS 511 & 214.895659 & 52.856515 &  3.21$^{+0.01}_{-0.01}$ & phot & 10.90$^{+0.08}_{-0.04}$ & 0.66$^{+0.16}_{-0.14}$ \\ 
 AEGIS 525 & 214.853928 & 52.861366 &  3.54$^{+0.03}_{-0.10}$ & phot & 11.37$^{+0.04}_{-0.04}$ & 0.81$^{+0.18}_{-0.16}$ \\ 
AEGIS  532 & 214.850584 & 52.866030 &  4.10$^{+0.05}_{-0.05}$ & phot & 10.82$^{+0.06}_{-0.03}$ & 0.96$^{+0.25}_{-0.22}$ \\ 
\enddata
\end{deluxetable*}

\vspace{-0.35in}
We performed an initial reduction of the NIRCam images in all four pointings, using version 1.5.3 of the \textit{JWST} Calibration Pipeline\footnote{\url{jwst-pipeline.readthedocs.io/en/latest/}} with some custom modifications.  We used the current (15 July 2022) set of NIRCam reference
files\footnote{\url{jwst-crds.stsci.edu}, jwst\_nircam\_0214.imap}, though 
we note that the majority were created pre-flight, including the flats and
photometric calibration references. We describe our reduction steps in greater detail in \citet{finkelstein22} and Bagley et al.~(\emph{in prep}).  Coadding the reduced observations into a single mosaic was performed using the drizzle algorithm with an inverse variance map weighting \citep{fruchter02,Casertano00} via the Resample step in the pipeline. The output mosaics have pixels scales of 0\farcs03/pixel.

Photometry was computed on PSF-matched images using SExtractor \citep{bertin96} v2.25.0 in two-image mode, with an inverse-variance weighted combination of the PSF-matched F277W and F356W images as the detection image.  Photometry was measured in all seven of the NIRCam bands observed by CEERS, as well as the F606W, F814W, F105W, F125W, F140W, and F160W \emph{HST} bands using data obtained by the CANDELS and 3D-HST surveys \citep{grogin11, koekemoer11,brammer12}.

\begin{figure*}[t]
\centering
\epsscale{1.15}
\plotone{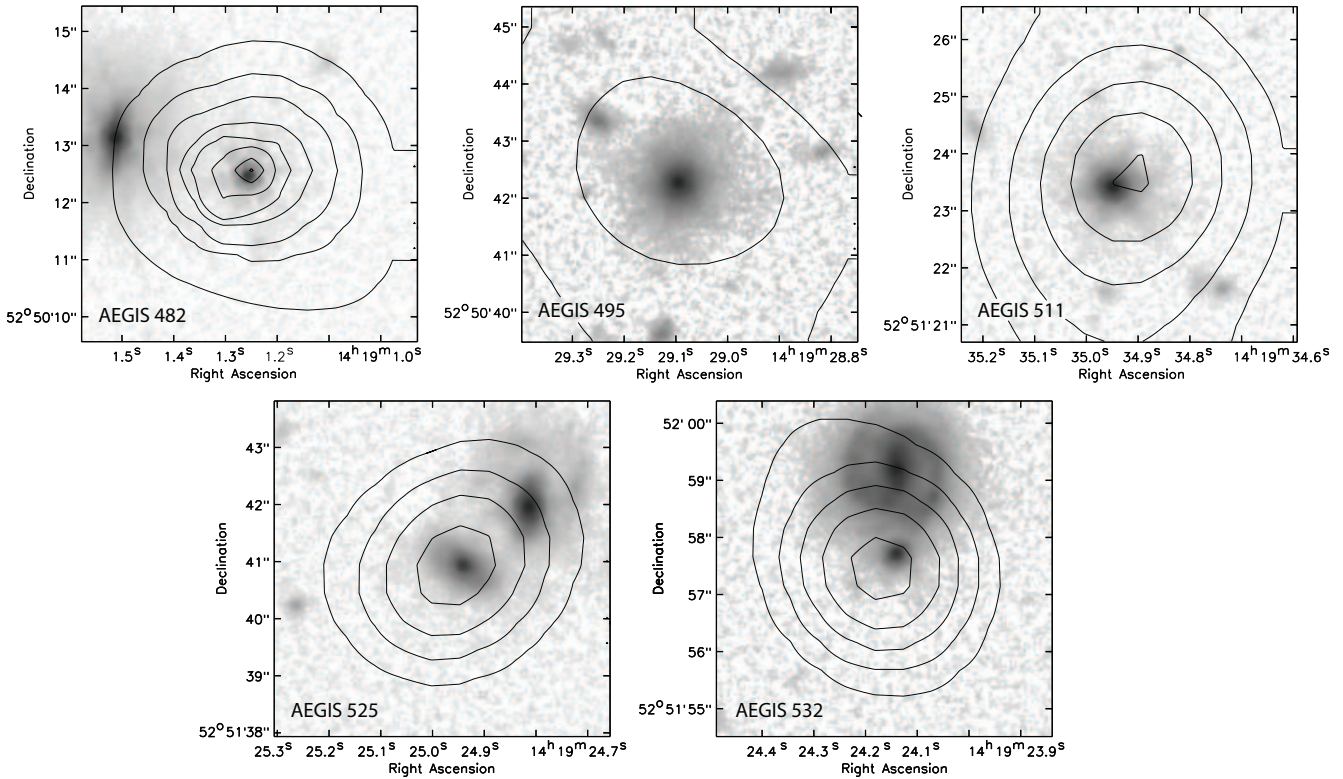}
\caption{X-ray emission contours overlaid on F356W images of the five AGN host galaxies with $3<z<5$ found in the first epoch of CEERS imaging. The X-ray emission is adaptively smoothed using the asmooth algorithm (Ebeling et al.~2006).  The contour levels begin at 1.5 times the local background and are logarithmically spaced thereafter.
\label{fig:xray_contours}}
\end{figure*}

\subsection{X-ray Observations}

The X-ray data used for this study come from the AEGIS-XD survey (\citealt{Nandra15}; hereafter N15), which consists of \emph{Chandra} ACIS-I observations with a characteristic exposure time of 800 ksec over all four of the epoch 1 CEERS pointings.  The survey has a flux limit of $1.5\times10^{-16}$ erg cm$^{-2}$ s$^{-1}$ in the 0.5-10 keV band, which corresponds to a luminosity limit that ranges from $5.3\times10^{42}$ erg s$^{-1}$ at $z=3$ to $1.7\times10^{43}$ erg s$^{-1}$ at $z=5$.

We make use of the published X-ray point-source catalog and counterpart associations presented in N15.  These associations were made by matching to a \emph{Spitzer}/IRAC 3.6 $\mu$m selected photometric catalog provided by the Rainbow Cosmological Surveys Database \citep{Barro11a, Barro11b}.  To take advantage of the increased spatial resolution provided by NIRCam, we independently cross-match the X-ray source list to our F277W+F356W-selected catalog using the maximum-likelihood technique described by \citet{ Sutherland_Saunders92}.  We find 48 X-ray sources matched to F277W+F356W counterparts in the CEERS imaging.

\begin{figure*}[t]
\centering
\epsscale{1.15}
\plotone{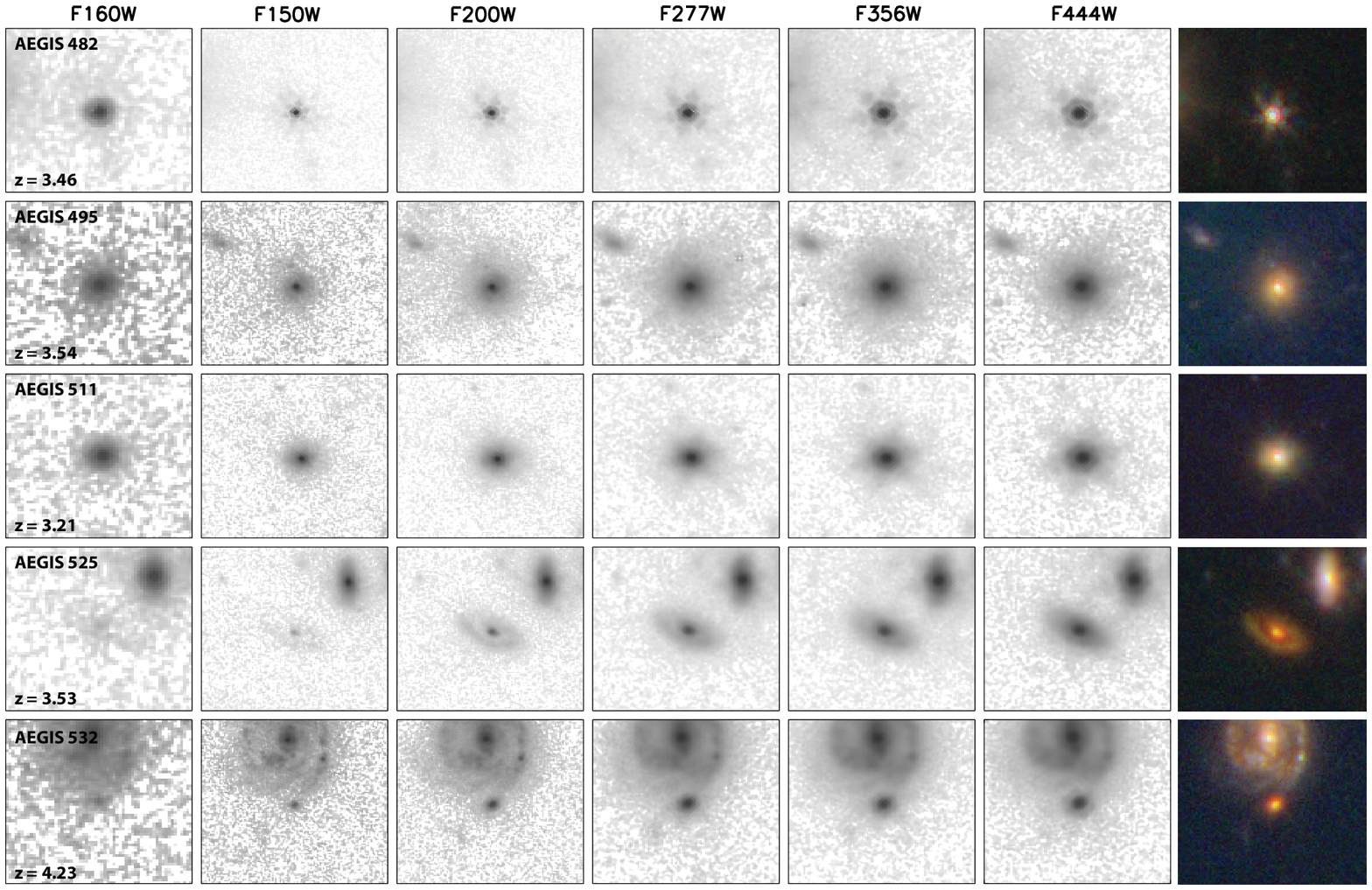}
\caption{Multiwavelength image cutouts of our primary AGN sample at $z>3$.  The images shown were taken in the F160W \emph{HST} WFC3 filter and five \emph{JWST} NIRCam short-wavelength (F150W, and F200W) and long-wavelength (F277W, F356W, and F444W) filters. The RGB images are composed of images in the F150W, F200W, and F277W filters. All images are $3\farcs5\times~3\farcs5$ in size.
\label{fig:mw_thumbs}}
\end{figure*}

\section{Sample Selection}

We use the EAzY \citep{brammer08} software to estimate the photometric redshifts of all sources in our multi-wavelength photometric catalog.  The spectral energy distribution (SED) fits were carried out using photometry in all seven of the NIRCam bands and all six of the \emph{HST} bands in our catalog.  The redshift range was allowed to vary from 0 to 12, and we used no zero-point corrections or luminosity priors. We chose the default template set \emph{tweak\_fsps\_QSF\_12\_v3} which was generated using the flexible stellar population synthesis (FSPS) code \citep{conroy09, conroy_gunn10}.  

Using a combination of our photometric redshifts and a compilation of published spectroscopic redshifts in the EGS field (N. Hathi 2022, private communication), we identify five X-ray detected AGN that have redshifts of $z>3$ in the current epoch of CEERS imaging.  These five AGN comprise our primary sample and are the focus of this study.  One of these sources, AEGIS 482, has a spectroscopic redshift of $z=3.465$ from the DEEP2 survey \citep{newman13}, while the remainder have photometric redshifts.  Information about each AGN in our sample is listed in Table 1. 

All but one X-ray source, AEGIS 532, have counterparts that agree with those published in N15.  AEGIS 532, however, was previously associated with a large foreground disk at $z=2.31$, but the X-ray emission is better centered on a neighboring galaxy that we estimate to be at $z=4.1$.  This can be seen in Figure \ref{fig:xray_contours}, which shows contours of the adaptively smoothed X-ray emission of each source overlaid on NIRCam imaging in the F356W band.  This source was not previously detected by the CANDELS survey and is blended with the foreground galaxy in the 3.6 $\mu$m \emph{Spitzer} IRAC imaging used for counterpart matching by N15.

The AGN in our $z>3$ sample have observed X-ray luminosities ranging from $1.8\times10^{43}$ erg s$^{-1}$ to $3.73\times10^{44}$ erg s$^{-1}$ in the 0.5-10 keV band (see Table 1), with an average luminosity of $1.27\times10^{44}$ erg s$^{-1}$.  Hardness ratios ($HR$) reported in N15 measured using the counts in the 0.5–2 and 2–7 keV bands show that AEGIS 482, AEGIS 495, and AEGIS 511 all have relatively unabsorbed emission ($HR <0$),  while AEGIS 525 and AEGIS 532 have harder emission indicative of a higher obscuring column density.  

We note that although the photometric redshifts from EAzY are calculated without accounting for AGN emission, we find only moderate AGN contribution to the galaxy SEDs of most of our sample (see \S5).  The only exception is AEGIS 482, but this source has a spectroscopically determined redshift.  Nonetheless, we checked the EAzY-derived redshifts of the five galaxies in our sample against those obtained from the X-CIGALE software \citep{yang20}, which performs SED fits using both AGN and stellar templates.  With the exception of AEGIS 482, which is our most X-ray luminous source, we find good general agreement between the two sets of redshifts.

\begin{figure*}[t]
\centering
\epsscale{1.15}
\plotone{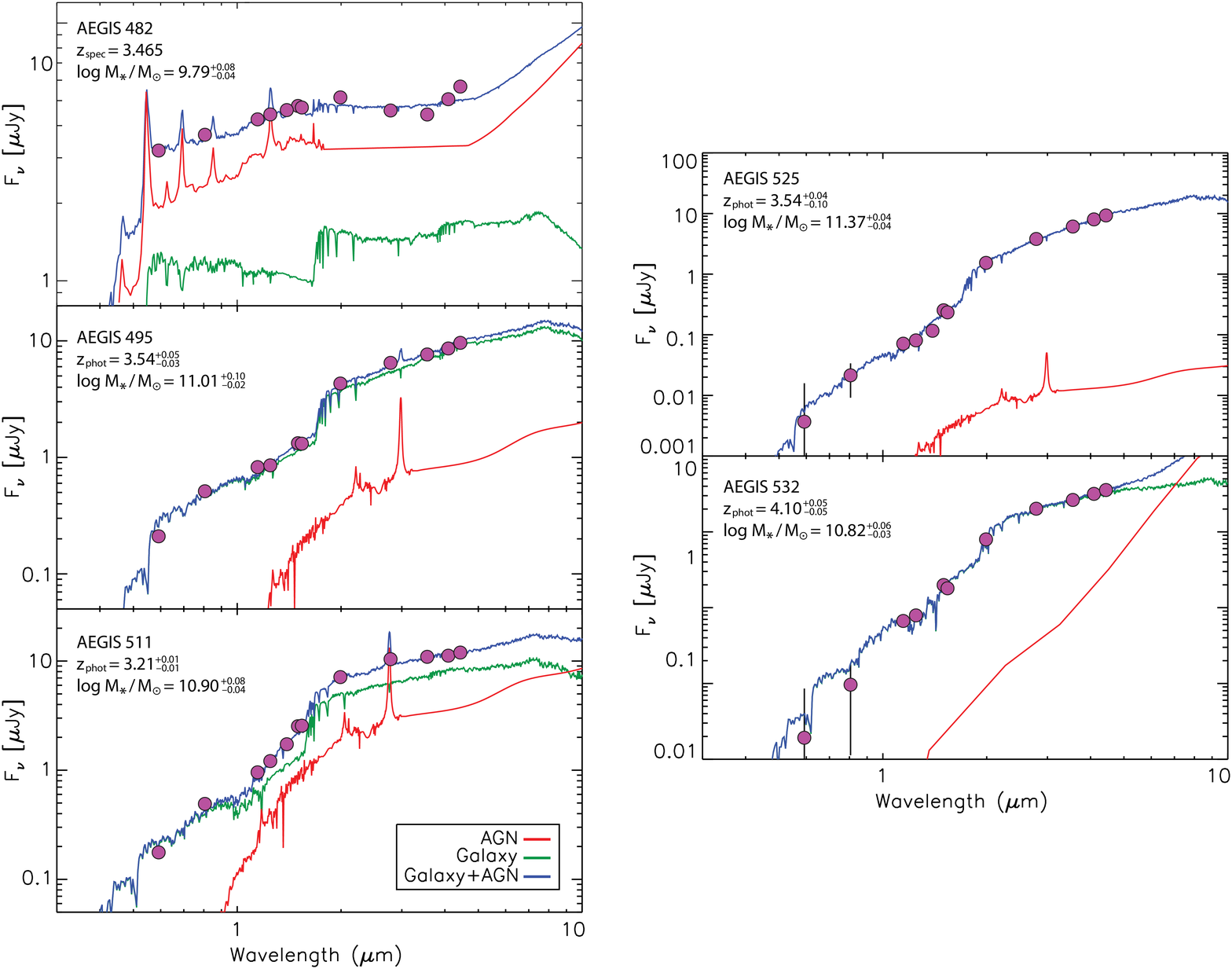}
\caption{Spectral energy distribution fits for our five AGN hosts at $z>3$.  Our measured photometric data from \emph{JWST} NIRCam, and \emph{HST} WFC3 and ACS are shown in magenta. The best-fit linear combination of each AGN template with each galaxy template in the full stellar population grid is shown in blue.  The galaxy and AGN contributions are shown separately in green and red, respectively. \label{fig:sed_plots}}
\end{figure*}

\section{Methodology}

Images of our AGN sample in five NIRCam bands can be seen in Figure \ref{fig:mw_thumbs}.  We assessed the morphology of their host galaxies through a combination of visual inspection and surface brightness profile fitting using the GALFIT software \citep{peng02}.  The visual classifications were carried out using the classification scheme presented in \citet{kocevski15}.  The GALFIT modeling was performed in the F356W band, which probes light redward of the Balmer break at the redshifts of all galaxies in the sample.  We provide GALFIT with empirical PSFs constructed from the four CEERS pointings and noise images that account for both the intrinsic image noise (e.g., background noise and readout noise) as well as added Poisson noise due to the objects themselves.  We fit each galaxy with a single S\'ersic profile \citep{sersic68}, but explore the need for additional components to help minimize the flux in our residual images.  Neighboring objects were fit simultaneously using single S\'ersic models.  

Stellar masses ($M_{*}$) and star formation rates (SFRs) were determined by performing SED fits using FAST v1.1 \citep{kriek19, aird18}, which allows for simultaneous fitting of both galaxy and AGN components.  For these fits we assume a \citet{chabier03} initial mass function, \citet{BC03} stellar population models, fixed solar metallicity, dust reddening of $A_V$ in the range 0–4 mag (assuming the \citealt{kriek_conroy13} dust attenuation curves), and “delayed-$\tau$” star formation histories with $\tau$ in the range 0.1–10 Gyr (e.g., \citealt{maraston10}).   We allow for an AGN component in the SED fit, adopting a library of eight empirically determined AGN templates (see Appendix A of \citealt{aird18}).  This provides a measure of various host properties (i.e., stellar mass, star formation rate, rest-frame colors) corrected for any contamination from non-stellar nuclear light.  Photometry in all of the \emph{JWST} and \emph{HST} bands discussed in \S2.1 were used for these fits.

\begin{figure}[t]
\centering
\epsscale{1.15}
\plotone{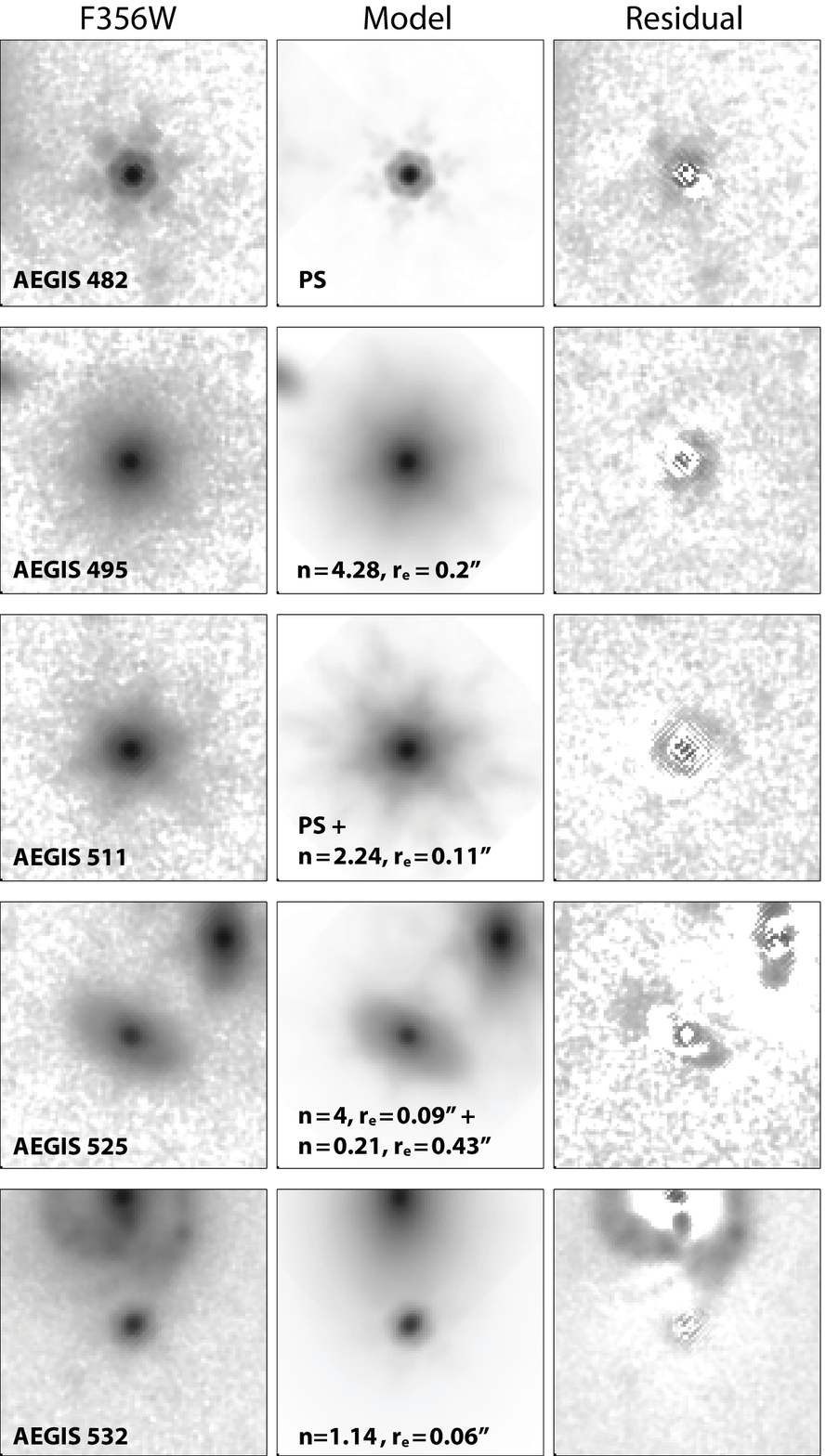}
\caption{Results of our two-dimensional surface brightness profile fitting. F356W images are shown in the left column, our best-fit GALFIT models are shown in the middle column, while the residuals (data-model) are shown in the right column. The best-fit S\'ersic index and effective radius for each source is listed. PS stands for point source. Images are $3^{\prime\prime}\times~3^{\prime\prime}$ in size. \label{fig:galfit_plot}}
\end{figure}

\section{Host Galaxy Properties}

The results of our two-component SED fits can be seen in Figure \ref{fig:sed_plots}.  We find that three of the five sources (AEGIS 495, AEGIS 525, and AEGIS 532) are fit with a moderate ($<10\%$) AGN contribution to the flux in the F356W band.  The non-stellar contribution rises to 31\% in AEGIS 511 and accounts for 70\% of the flux from AEGIS 482, whose SED is best fit using a Type-1 QSO template.  The former shows signs of diffraction spikes in our long wavelength images, consistent with a nuclear point source, while the latter is dominated by point-like emission in both the \emph{JWST} and \emph{HST} imaging (see \S4.1 below).  Furthermore, the spectrum of AEGIS 482 from the DEEP2 survey shows broad \CIII~and \CIV~emission lines, consistent with the QSO SED fit to this source.

We find that the four well-resolved hosts have stellar masses in the range of ${\rm log}(M_{*}/{\rm M_{\odot}} )= 10.82-11.37$ (see Table 1), making them among the most massive galaxies detected at this redshift range in the current CEERS pointings (Finkelstein et al., \emph{in prep}), even after accounting for nuclear light from the AGN . In the following sections, we examine the morphologies and star-formation activity of these galaxies in greater detail.

\subsection{Morphologies}

Based on our visual classifications, three of the AGN (AEGIS 495, AEGIS 511 and AEGIS 532) have spheroidal hosts, while one (AEGIS 525) is found in a disk galaxy with a prominent bulge component at longer wavelengths. The final source (AEGIS 482) appears point-like in all seven NIRCam filters observed by CEERS.  Our SED fit for AEGIS 525, the single disk in the sample, finds that minimal AGN light is needed in the reddest NIRCam bandpasses, suggesting the central bulge component is indeed a stellar bulge and not nuclear emission from the AGN.  Finally, none of the AGN hosts show strong asymmetries or distortions that might indicate a recent interaction or merger event. All of them are visually classified as undisturbed.  The close companions to AEGIS 525 and AEGIS 532 are both foreground galaxies seen in projection, with redshifts of 1.35 and 2.07, respectively. 

\begin{figure*}[t]
\centering
\epsscale{1.0}
\plotone{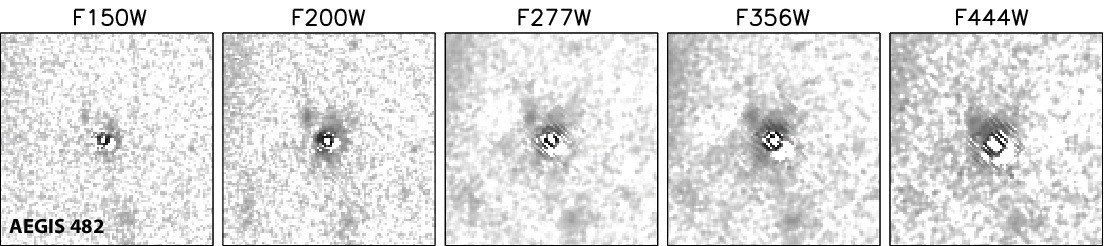}
\caption{Residual images of AEGIS 482 after subtracting the best-fit point-source model in five NIRCam bands. Images are $3^{\prime\prime}\times~3^{\prime\prime}$ in size. \label{fig:aegis482_res}}
\end{figure*}

The results of our two-dimensional surface brightness profile fitting can be seen in Figure \ref{fig:galfit_plot}.  The only source not fit with a S\'ersic profile is AEGIS 482, which we find is best fit using a point-source model.  In Figure \ref{fig:aegis482_res} we show the residual images of these fits in five NIRCam bands ranging from F150W to F444W.  While point-like emission dominates the light from this source, we see extended structure that is visible in multiple bands.  The underlying host appears particularly elongated at F200W.  We also note a possible nearby companion or stellar clump (e.g., \citealt{forster_schreiber11}) in the residual images.  The companion is most clearly discernible in F200W, but is present in all five bands.

The hosts of AEGIS 495 and 511 are fit with S\'ersic indices ($n$) of $n=4.28\pm0.03$ and $n=3.49\pm0.01$, respectively, in agreement with our visual classification of these galaxies as spheroids.  However, since the SED of AEGIS 511 was fit with moderate AGN contribution, we also modeled it using a point-source component in addition to a S\'ersic profile.  The resulting fit attributes 17.6\% of the light in F356W to the unresolved nuclear component, which reduces the resulting S\'ersic index to $n=2.24\pm0.02$.

\begin{figure}[t]
\centering
\vspace{0.15in}
\epsscale{1.15}
\plotone{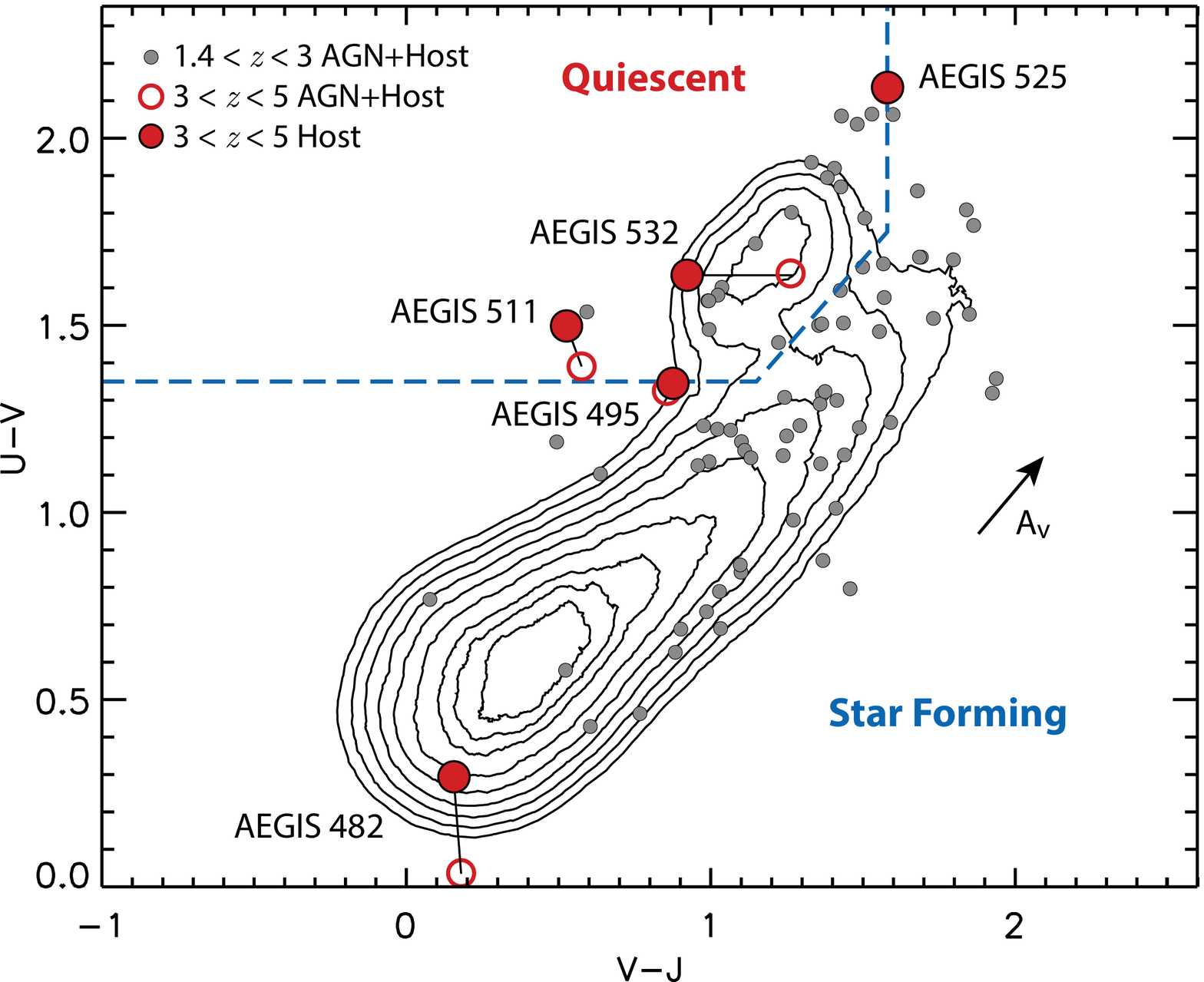}
\caption{UVJ color diagram.  Contours and grey circles show the color distribution of galaxies and the hosts of X-ray selected AGN, respectively, in the redshift range $1.4 < z < 3.0$ in the CANDELS fields. The blue dashed line denotes the dividing line used to separate quiescent and star-forming galaxies. Red open circles show the rest-frame colors of the AGN+galaxy emission of our $3<z<5$ sample, while red filled circles are colors corrected for the AGN emission predicted in each band from our two-component SED fits. \label{fig:uvj}}
\end{figure}

A single S\'ersic fit of AEGIS 525 results in $n=4.20+/-0.04$, indicative of the prominent bulge in this system.  We also performed a double S\'ersic fit with one component fixed to a $n=4$ de Vaucouleurs profile \citep{deVauc48}.  In this case, the second component is fit with a large effective radius ($r_{e} = 0\farcs43 \pm 0\farcs01$) and a relatively flat surface brightness profile ($n=0.21\pm 0.01$). The resulting bulge-to-total ratio of the galaxy is 0.69, consistent with a bulge-dominated disk. 

The host of AEGIS 532, the most compact galaxy in the sample, is best fit with a small effective radius of $r_{e} = 0\farcs06 \pm 0\farcs01$ and a S\'ersic index of $n=1.14\pm 0.05$.  This raises the possibility that it is a massive, compact disk such as those reported at $z\sim2$ (e.g., \citealt{weinzirl11, vanderWel11}).  However, we caution that this fit may be impacted by the large, nearby galaxy whose spiral structure is poorly fit with a single S\'ersic profile.

\subsection{Star-Formation Activity}

In Figure \ref{fig:uvj} we show the rest-frame $V-J$ versus $U-V$ colors of the five AGN hosts, both with and without correcting for AGN emission.  The colors were determined by convolving the redshifted $U$, $V$, and $J$ filter bandpasses with the best-fitting templates returned by FAST for each source.  The green circles show the color of the AGN+galaxy emission, while red circles denote galaxy-only emission corrected for any AGN contamination.  The dashed blue line denotes the color cut of \citet{williams09} for identifying quiescent galaxies.  

We find AEGIS 511, AEGIS 525 and AEGIS 532 are all located within the quiescent boundary. AEGIS 525 lies near the red border of the quiescent region and is offset from the locus of quiescent galaxies observed at $z\sim2$, which likely reflects additional dust reddening in this galaxy.  Our SED modeling returns a best-fit dust attenuation ($A_{V}$) of 1.9 for this source, consistent with its redder $UVJ$ color.

AEGIS 495 sits just blueward of the lower $U-V$ boundary.  This region is notable in that the bluer $U-V$ colors are consistent with young quiescent galaxies, such as those that have recently experienced a burst of star-formation followed by a rapid truncation of activity (e.g., \citealt{whitaker12, Schreiber18}).  It suggests this galaxy may be recently quenched or in a post-starburst phase.  Finally, the best-fit galaxy template for AEGIS 482 places it squarely within the locus of unobscured, star-forming galaxies.


To further assess the star-formation activity of our sample, we compared their specific star formation rates (sSFR) to that of the star-forming main sequence (e.g., Noeske et al.~2007) at $z=3.5$, which is found to have sSFR$_{\rm MS}$ = 1.5 Gyr$^{-1}$ \citep{Schreiber17, Schreiber18}.  We find that the hosts of AEGIS 511, AEGIS 525, and AEGIS 532 all have sSFR that are a factor of ten or more below that of the main sequence, while AEGIS 495 is suppressed by a factor of 2.9, in general agreement with their $UVJ$ colors.

Recently, \citet{Carnall22} reported that the hosts of both AEGIS 525 and AEGIS 532 are quiescent.  That study used the time-dependent quiescent selection criteria ${\rm sSFR} < 0.2 / t_{obs}$, where sSFR is the star-formation rate per unit mass and $t_{obs}$ is the age of the Universe at the redshift of the galaxy.   We find that AEGIS 511, AEGIS 525, and AEGIS 532 would all be considered quiescent using this criteria, while AEGIS 495 lies just above the threshold.  This galaxy is also the source that falls in the post-starburst region of the $UVJ$ diagram. All this together suggests the host of AEGIS 495 is the sole well-resolved galaxy in our sample with a moderate level of ongoing star formation, albeit potentially suppressed relative to that of the star-forming main sequence.

\section{Discussion \& Conclusions}

We examine the rest-frame optical host properties of five X-ray luminous AGN detected at $3 < z < 5$ using \emph{JWST} NIRCam imaging taken as part of the CEERS program.  Four of the hosts are spatially well resolved and a visual assessment of their morphologies reveals three are spheroidal systems and one a bulge-dominated disk.  None of the galaxies show strong morphological disturbances indicative of a recent interaction or merger event, however point-source subtraction reveals a potential close companion near the most X-ray luminous AGN in the sample. We find the four resolved hosts have rest-frame colors that place them in the quiescent and post-starburst regions of the $UVJ$ diagram.  We also perform two-component SED fits using both galaxy and AGN templates and find that the sSFR of all four galaxies are at least a factor of $\sim3$ below the star-forming main sequence at z=3.5, confirming that they have recently quenched or are potentially in the process of quenching.

The small sample size notwithstanding, the properties of these galaxies stand in contrast to those measured among the bulk of the AGN population at $z\sim2$, where hosts are predominately normal, star-forming systems with a large disk fraction (e.g., \citealt{Schawinski11, Kocevski12, Mullaney12, Rosario13, florez20, Ji22}).  However, AGN activity is not uncommon among the quiescent population at these redshifts.  Studies find 20-25\% of massive ($M_{*} > 10^{10} M_{\odot}$) quiescent galaxies host an X-ray AGN at $z\sim2$ \citep{Kocevski17, Wang17}; a five-fold increase over the fraction measured at $z\sim1$ (e.g., \citealt{georgakakis08}).  This fraction increases further at higher redshifts, where \citet{Schreiber18} report that 33\% of their sample of young massive quiescent galaxies at $3 < z < 4$ are X-ray detected.  
This strong evolution in AGN fraction implies a much higher AGN duty cycle in passive galaxies at these redshifts.  Coupled with the host properties that we observe, this suggests distant quiescent galaxies have increased residual cold-gas reservoirs that can continue to fuel SMBH growth even after star formation has been curtailed.

 The ubiquity of AGN in massive quiescent galaxies at $z>3$ is notable because energy injection from AGN has been used widely in semi-analytic models and cosmological simulations as a key quenching mechanism.  The need for strong feedback is particularly acute for massive quiescent galaxies at the redshifts of our sample given the limited time available (1-2 Gyr) to form them and fully shut down their star formation activity.  
 
 If these host galaxies were rapidly quenched as described by the radiative (or quasar) mode feedback model, where AGN-driven winds help to remove a galaxy's cold-gas supply, then we would be observing them in the post blow-out phase, when the obscuring column density has dropped enough for the AGN to be visible and the star-formation and nuclear activity are in gradual decline. A galaxy in a similar state was recently reported by \citet{kubo22}, who detect a Type 2 QSO in a massive ($M_{*} = 10^{11.3}$ M$_{\odot}$) quiescent galaxy at $z=3.09$.  In that case, strong ionized gas outflows are detected via broad [OIII] emission, providing direct evidence of energy injection by the AGN.   Follow-up observations will be needed to confirm if similar outflows are present in our sample, but this example demonstrates a plausible scenario to explain the host properties that we observe.
 
Of course, one proposed mechanism for triggering quasar mode feedback is major galaxy-galaxy mergers, which are thought to prompt the radiatively efficient accretion that ultimately powers the quenching outflows.  With the exception of AEGIS 482, which shows signs of a possible companion in the GALFIT residual images, our hosts do not exhibit the strong morphological disturbances that might be expected if these systems experienced a recent major merger event.  A delay between the merger and the onset of AGN activity that is longer than the relaxation time of a galaxy (typically a few hundred Myr; \citealt{Lotz10}), coupled with surface brightness dimming, might cause the most obvious merger signatures to fade below our detection limit.  Such time lags are expected \citep{DiMatteo05, Springel05a} and are often invoked as a possible explanation for the lack of merger signatures in AGN hosts (e.g., \citealt{Cisternas11, Kocevski12}).  However, AEGIS 525, which has a significant disk component, is unlikely to have experienced a major disruptive merger in the recent past.  Although disks can reform following a gas-rich merger \citep{robertson06, bundy10}, the timescale required ($\sim1$ Gyr) make this unlikely given the redshift of the source.

It should be noted that for moderate luminosity AGN, like many in our sample, the implied black hole mass accretion rate is modest (a few ${\rm M_{\odot}}$ yr$^{-1}$ for $L_{\rm bol}/L_{X}\sim30$ and a radiative efficiency of 0.1) and could be sustained over several duty cycles with only modest amounts of gas ($\sim10^{9}~{\rm M_{\odot}}$). If such gas reservoirs, which are common in the circumnuclear region of local spiral galaxies, are present in the circumnuclear region of $z>3$ galaxies, they can readily fuel such AGN without the need for large-scale gas transport by mergers, as long as local circumnuclear processes can drain angular momentum from the gas and drive it down to the black hole accretion disk (e.g., \citealt{jogee06, Hopkins14}).

Alternative triggering mechanisms include minor mergers or rapid cold flow accretion \citep{dekel09}, which can funnel gas to the centers of galaxies on short timescales \citep{bournaud11}.  According to recent studies using cosmological hydrodynamic simulations, these mechanisms may in fact be responsible for triggering the bulk of AGN activity at early times rather than major merger \citep{steinborn18, sharma21}.

Whether the initial quenching of these galaxies is ultimately due to AGN feedback or another cause, such as simple rapid gas exhaustion, remains to be determined.  However, the presence of luminous AGN in these systems, and the observed high duty cycle in massive quiescent galaxies at similar redshifts, implies AGN can input a significant amount of energy into their hosts after star formation has ceased, which may heat the halos of these systems and prevent renewed star formation.  A larger sample size will be needed to determine if this is a common role that moderate-luminosity AGN play during the era of galaxy assembly.  In a future paper, we plan to expand our sample size using the full CEERS dataset and do a detailed comparison of the host properties to a mass and redshift matched control sample of inactive (non-AGN) galaxies.

\section{Acknowledgments}

This work is supported by NASA grants JWST-ERS-01345 and JWST-AR-02446 and based on observations made with the NASA/ESA/CSA James Webb Space Telescope. The data were obtained from the Mikulski Archive for Space Telescopes at the Space Telescope Science Institute, which is operated by the Association of Universities for Research in Astronomy, Inc., under NASA contract NAS 5-03127 for JWST.  This work also made use of the Rainbow Cosmological Surveys Database, which is operated by the Centro de Astrobiología (CAB/INTA), partnered with the University of California Observatories at Santa Cruz (UCO/Lick,UCSC).


\end{document}